\documentclass[prb,preprint]{revtex4-1} 

\usepackage{amsmath}  
\usepackage{amsfonts} 
\usepackage{graphicx} 

\usepackage{braket}
\usepackage{tikz}
\usetikzlibrary{arrows.meta}

\begin{document}


\title{A quantum treatment of the Stern-Gerlach experiment}

\author{K. M. Fonseca-Romero}
\email{kmfonsecar@unal.edu.co} 
\affiliation{Departamento de Física, Universidad Nacional de Colombia - Sede Bogotá, Facultad de Ciencias, Grupo de Óptica e Información Cuántica, Carrera 30 Calle 45-03, C.P. 111321, Bogotá, Colombia}


\date{\today}

\begin{abstract}
Most textbooks introduce the concept of spin by presenting the Stern-Gerlach experiment with the aid of Newtonian atomic trajectories. 
However, to understand how both spatial and spin degrees of freedom evolve over time and how the latter influence experimental outcomes, it is essential to employ a quantum approach.
In this paper, we offer two simple methods, the Baker-Campbell-Hausdorff formula and the direct integration of the Schrödinger equation in an interaction picture, to determine the corresponding evolution operator. 
We not only provide an interpretation of the individual terms within this operator but also establish connections with semiclassical calculations, when feasible. 
Moreover, we compute the wave function and  touch upon the concept of position-spin entanglement to illustrate how a full quantum description of the Stern-Gerlach experiment can open doors to topics like quantum measurement and nonlocality.

\end{abstract}



\maketitle 

\section{Introduction} 

The Stern-Gerlach experiment (SGE) serves as a significant historical milestone in physics. 
Initially, it demonstrated the concept of space quantization~\cite{spacequantization} and, in retrospect, marked the first measurement of electron spin. 
This classic experiment has a conventional role in physics education in teaching the quantum formalism~\cite{zhu2011improving}.
The SGE's potential for exploration goes beyond concepts like state preparation, time evolution, and distinction between Hilbert space and physical space.
It can be employed to delve into a variety of quantum concepts, including the examination of Hilbert spaces for composite systems, the study of mixed quantum states, and the exploration of quantum measurement processes, entanglement, and  nonlocality.
To fully realize this rich potential for exploration, a comprehensive quantum treatment, often missing from standard textbooks, is essential. 
Quantum descriptions of the SGE have been previously employed, for example, in references~\onlinecite{platt1990modern} and \onlinecite{rodriguez2017full}.
Platt~\cite{platt1990modern} wrote the Schrödinger equation for the SGE and justified an effective Hamiltonian, commonly used to describe this experiment with a single-component magnetic field. Additionally, Rodríguez \textit{et al.}\cite{rodriguez2017full} computed the time-evolution operator\cite{robinett1996quantum,balasubramanian2001time} corresponding to this effective Hamiltonian, using the Norman-Wei~\cite{wei1963lie,wei1964global} factorization method.
Here, we present two simpler methods to obtain the evolution operator for the effective Hamiltonian and arbitrary spin, as a product of simple unitary operators.
We also interpret each of these operators, compute the evolution of an initial state with spatial Gaussian dependence, estimate the beam splitting of the original experiment, and briefly discuss position-spin entanglement in a particular example.

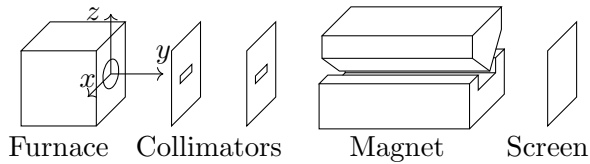
\begin{figure}[h!]
\centering
\begin{tikzpicture}[scale=1]
   \draw (0,1,0) -- (1,1,0) -- (1,1,1) -- (0,1,1) -- cycle;
   \draw (0,1,1) -- (0,0,1) -- (1,0,1) -- (1,1,1);  
   \draw (1,0,1) -- (1,0,0) -- (1,1,0);
   \draw[color=black,rotate around={-10:(1,0.5,0.5)}] (1,0.5,0.5) ellipse [x radius=0.1, y radius=0.2];
   \draw (2,0,1) -- (2,0,0) -- (2,1,0) -- (2,1,1) -- cycle;
   \draw (2,0.45,0.7) -- (2,0.45,0.3) -- (2,0.55,0.3) -- (2,0.55,0.7) -- cycle;
   \draw (3,0,1) -- (3,0,0) -- (3,1,0) -- (3,1,1) -- cycle;   
   \draw (3,0.45,0.7) -- (3,0.45,0.3) -- (3,0.55,0.3) -- (3,0.55,0.7) -- cycle;
   \draw (4,1.2,1) -- (4,1.2,0) -- (6,1.2,0) -- (6,1.2,1) -- cycle;   
   \draw (4,1.2,1) -- (4,0.9,1) -- (6,0.9,1);
   \draw (6,1.2,1) -- (6,0.9,1) -- (6,0.5,0.4) -- (6,0.9,0) -- (6,1.2,0);
   \draw (4,0.9,1) -- (4,0.5,0.4) -- (6,0.5,0.4);
   \draw (4,0.6,1.1) -- (6,0.6,1.1) -- (6,0.6,0.8) -- (4,0.6,0.8) -- cycle;
   \draw (4,0.6,1.1) -- (4,0,1.1) -- (6,0,1.1) -- (6,0.6,1.1);
   \draw (6,0.6,0.8) -- (6,0.4,0.8) -- (6,0.4,0.2) -- (6,0.6,0.2) -- (6,0.6,-0.1) -- (6,0,-0.1) -- (6,0,1.1);
   \draw (5.965,0.6,0.2) -- (6,0.6,0.2);
   \draw (5.885,0.6,-0.1) -- (6,0.6,-0.1);
   \draw (4,0.42,0.2) -- (4,0.4,0.2) -- (6,0.4,0.2);
   \draw (4,0.4,0.29) -- (4,0.4,0.2);
    \draw (7,0,1) -- (7,0,0) -- (7,1,0) -- (7,1,1) -- cycle;   
   \draw[->] (1,0.5,0.5) -- (1.7,0.5,0.5) node[above] {$y$};
   \draw[->] (1,0.5,0.5) -- (1,1.3,0.5) node[left] {$z$};
   \draw[->] (1,0.5,0.5) -- (1,0.5,1.3) node[above] {$x$};
   \node[below] at (0.5,0,1) {Furnace};
   \node[below] at (2.5,0,1) {Collimators};
   \node[below] at (5,0,1) {Magnet};
   \node[below] at (7,0,1) {Screen};
\end{tikzpicture}
\caption{Stern-Gerlach experiment}
\label{fig:SGE}
\end{figure}

Figure \ref{fig:SGE} shows the schematics of the SGE~\cite{friedrich1998space}.
A beam of silver atoms leaving a furnace at about $v = 660$ m/s is collimated by narrow ($2a = 0.03$ mm) slits.
The atoms traverse a 3.5 cm ($L$) magnet which produces a $B_0 = 0.1$ T magnetic field with a vertical ($z$) gradient of $\beta = 10$ T/cm.
Then, the atoms are deposited on a glass plate.
The second collimator and the glass screen are assumed to be in contact with the magnet.

We consider the dynamics of the atoms while traversing the magnet.
The interaction of the atoms, considered as neutral point particles with magnetic dipolar moment, with the magnetic field is modeled by the Hamiltonian
\begin{equation}
 \hat{H} = \frac{\hat{\boldsymbol{p}}^2}{2M} - \gamma \hat{\boldsymbol{S}} \cdot \boldsymbol{B}(\hat{\boldsymbol{r}}),
\end{equation}
where $\hat{\boldsymbol{p}}$ , $\hat{\boldsymbol{r}}$, and $\hat{\boldsymbol{S}}$ are, respectively, the momentum, position and spin operators, and $\boldsymbol{B}(\boldsymbol{r})$ is the magnetic field at position $\boldsymbol{r}.$
Here, $M$ is the mass of the atom, and $\gamma = -g \mu_B/\hbar,$ where $g$ is the gyromagnetic ratio and $\mu_B$ is the Bohr magneton.
The magnetic field is assumed to be
\begin{equation}
 \boldsymbol{B}(\boldsymbol{r}) = -\beta x \hat{\boldsymbol{\imath}} + (B_0 + \beta z) \hat{\boldsymbol{k}}.
\end{equation}
If $B_0 \gg |B_x|$, $\hat{S}_x$ precesses rapidly around the $z$-axis~\cite{platt1990modern,hannout1998quantum}, such that the interaction with the $B_x$ component averages out.
Hence, in the limit of large $B_0$, the effective Hamiltonian $\hat{H}_{\textrm{e}}$  contains only the $z$-component,
\begin{equation}
  \hat{H}_{\textrm{e}} = \frac{\hat{\boldsymbol{p}}^2}{2M} - \gamma (B_0 + \beta \hat{z}) \hat{S}_z 
  = \hat{H}_2 + \hat{H}_1,
\end{equation}
where $\hat{H}_2$ is the kinetic Hamiltonian $\hat{\boldsymbol{p}}^2/(2M).$
The corresponding evolution operator, $U_{\textrm{e}}(t)$, satisfies the Schrödinger equation
\begin{equation}
 i\hbar \frac{d\hat{U}_{\textrm{e}}(t)}{dt} = \hat{H}_{\textrm{e}} \hat{U}_{\textrm{e}}(t),
\end{equation}
subject to the initial condition $\hat{U}_{\textrm{e}}(t=0) = \hat{I},$ where $\hat{I}$ is the identity operator.
The state of the system at time $t,$ is $\ket{\Psi(t)} = \hat{U}_{\textrm{e}}(t) \ket{\Psi(0)} = \exp (-i  \hat{H}_{\textrm{e}} t/\hbar) \ket{\Psi(0)}.$
Expanding the evolution operator as a Taylor series in terms of $t$ often proves challenging for calculating the evolved state, mainly due to the presence of non-commuting operators in the Hamiltonian. 
Consequently, various approaches, such as the factorization method and the utilization of operator identities like the BCH formula, have been devised to address this difficulty.

The evolution operator $\hat{U}_{\textrm{e}}(t)$ was determined in Ref. \onlinecite{rodriguez2017full} through the application of the Norman-Wei factorization method~\cite{wei1963lie,wei1964global}. 
However, this method, though effective, is rather lengthy and may not be the most suitable choice for classroom discussions.
In this work, we present two more straightforward alternatives for obtaining the evolution operator. 
The first method is a more direct approach, utilizing a specific instance~\cite{robinett1996quantum} of the Baker-Campbell-Hausdorff formula, while the second employs an intermediate interaction picture.

If $\hat{A}$ and $\hat{B}$ are operators which commute with their double commutators $[\hat{A},[\hat{A},\hat{B}]]$ and $[\hat{B},[\hat{B},\hat{A}]],$ then
\begin{equation} \label{eq:CBH}
 e^{\hat{A}} = 
 e^{\hat{B}} e^{\hat{A}-\hat{B}+\frac{[\hat{A},\hat{B}]}{2} + \frac{1}{12} ( [\hat{B},[\hat{B},\hat{A}]] -[\hat{A},[\hat{A},\hat{B}]] )}.
\end{equation}
If  $\hat{A} = -i \hat{H}_{\textrm{e}}t/\hbar$ and $\hat{B} =-i \hat{H}_1 t/\hbar= i\gamma t (B_0+\beta \hat{z})\hat{S}_z/\hbar,$ using the canonical commutators of position and momentum we get
\begin{align}
 \hat{A}-\hat{B} &= -i \frac{\hat{H}_2 t}{\hbar}=  \frac{-i\hat{\boldsymbol{p}}^2 t}{2\hbar M}, \\
[ \hat{A}, \hat{B} ]  & = \frac{\gamma t^2}{\hbar^2} [\hat{H}_{\textrm{e}}, \hat{H}_1]  = -\frac{i\gamma\beta t^2}{\hbar M}  \hat{p}_z \hat{S}_z,\\
[ \hat{A},[ \hat{A}, \hat{B} ] ]&  
= i  \frac{\gamma^2\beta^2 t^3}{\hbar} \hat{S}^2_z = -[\hat{B},[\hat{B},\hat{A}]].
\end{align}
Since all terms of the second term of the right hand side of \eqref{eq:CBH} commute among themselves, we can cast the evolution operator in the form
\begin{equation}
 \hat{U}_{\textrm{e}}(t) = \hat{U}_1(t) \hat{U}_{2a}(t)  \hat{U}_{2b}(t)  \hat{U}_{2c}(t),
\end{equation}
where
\begin{align} \label{eq:U1}
  \hat{U}_1(t) & = e^{\frac{i\gamma t}{\hbar} (B_0+\beta \hat{z})\hat{S}_z},\\
  \hat{U}_{2a}(t) & =e^{-\frac{i \hat{\boldsymbol{p}}^2 t}{2M\hbar}} ,\\
  \hat{U}_{2b}(t) & =e^{ \frac{-i\gamma \beta t^2}{2\hbar M} \hat{S}_z \hat{p}_z },\\
   \hat{U}_{2c}(t) & =e^{-\frac{i\gamma^2 \beta^2 t^3}{6\hbar M} \hat{S}_z^2}.
\end{align}

For the second method, we assume that the evolution operator is written as $\hat{U}=\hat{U}_1\hat{U}_2$, and that $\hat{H} = \hat{H}_1 + \hat{H}_2$, and $\hat{U}_1$ is the evolution operator corresponding to $\hat{H}_1$.
Under these circumstances, the Schrödinger equation
\begin{equation}
 i\hbar \frac{d(\hat{U}_1\hat{U}_2)}{dt} = (\hat{H}_1+\hat{H}_2) \hat{U}_1\hat{U}_2,  
\end{equation}
leads to the equation of motion for $\hat{U}_2$,
\begin{equation}
  i\hbar \frac{d\hat{U}_2}{dt} = \hat{U}_1^\dagger \hat{H}_2 \hat{U}_1\hat{U}_2 = \tilde{H}_2(t) \hat{U}_2.
\end{equation}
That is, $\hat{U}_2$ is the evolution operator of the new Hamiltonian $ \tilde{H}_2(t) =  \hat{U}_1^\dagger \hat{H}_2 \hat{U}_1.$

We choose $\hat{H}_1 = -\gamma(B_0+\beta \hat{z})\hat{S}_z, $ such that $ \hat{U}_1(t)$ is given by eq.  \eqref{eq:U1}. 
The new Hamiltonian, in the interaction picture, is
\begin{equation}
  \tilde{H}_2(t) = \hat{U}_1^\dagger (t)  \frac{\hat{\boldsymbol{p}}^2  }{2M} \hat{U}_1(t)
  = \frac{\hat{p}_x^2 + \hat{p}_y^2 }{2M} + \frac{\left(\hat{U}_1^\dagger (t)  \hat{p}_z \hat{U}_1 (t) \right)^2 }{2M},
\end{equation}
because $\hat{z}$ and $\hat{S}_z$ commute with the momentum operators $\hat{p}_x$ and $\hat{p}_y.$ We also use $\hat{U}_1^\dagger (t)  \hat{p}_z^2 \hat{U}_1 (t)  =  \hat{U}_1^\dagger (t)  \hat{p}_z \hat{U}_1 (t) \hat{U}_1^\dagger (t)  \hat{p}_z \hat{U}_1 (t).$ 

Using the easily demonstrated (see Appendix) Cox operator expansion~\cite{wilcox1967exponential}
\begin{equation} \label{eq:expansion}
 e^{x \hat{A}} \hat{B}  e^{x \hat{A}} = \hat{B} +\frac{x}{1!}[\hat{A},\hat{B}] +\frac{x^2}{2!}[\hat{A},[\hat{A},\hat{B}]] + \cdots
\end{equation}
and the canonical commutation relations of position and momentum, we obtain 
\begin{equation} 
\hat{U}_1^\dagger (t)  \hat{p}_z  \hat{U} (t)  
= \hat{p}_z + \gamma \beta t \hat{S}_z.
\end{equation} 
All terms beyond the first of the expansion \eqref{eq:expansion} vanish because $\hat{z}$ commutes with $\hat{S}_z.$
We write the Schr\"odinger equation for $\hat{U}_2$ as
\begin{eqnarray}
 i\hbar \frac{d\hat{U}_2}{dt} &=& \left( \frac{\hat{p}_x^2 + \hat{p}_y^2 }{2M} + \frac{\left(  \hat{p}_z + \gamma \beta t \hat{S}_z \right)^2 }{2M} \right) \hat{U}_2(t)\\
&=& \left(  \frac{\hat{\boldsymbol{p}}^2}{2M} + \frac{\gamma \beta t \hat{S}_z \hat{p}_z}{M} + \frac{\gamma^2 \beta^2 t^2 \hat{S}_z^2}{2M}\right) \hat{U}_2(t).
\end{eqnarray}
Taking into account that $[\tilde{H}_2(t_1),\tilde{H}_2(t_2)]=0$ we can immediately integrate this equation
\begin{equation}
\hat{U}_2(t) = e^{ -i\frac{\hat{\boldsymbol{p}}^2 t}{2M\hbar}  -i \frac{\gamma \beta t^2 \hat{S}_z \hat{p}_z}{2\hbar M} -i \frac{\gamma^2 \beta^2 t^3 \hat{S}_z^2}{6\hbar M}}.
\end{equation}
Since all operators in the exponential commute with each other, we can separate each contribution, in an arbitrary order and recover the result obtained with the first method. 

The operator $ \hat{U}_{2c}(t)= \exp(-{i\gamma^2 \beta^2 t^3}\hat{S}_z^2/({6\hbar M} ))$ describes a nonlinear transformation of the spin, which reduces to a global phase for spin 1/2 systems (like the original SGE).
For example,  the $x$ component of the spin, in the Heisenberg picture, is transformed as $\hat{S}_x(t) = \hat{U}_{2c}^\dagger (t) \hat{S}_x \hat{U}_{2c}(t).$
For a spin one system,  $\hat{U}_{2c}(t)$ is of the form $e^{i\alpha} \hat{S}_z^2,$ where $\alpha =  - \gamma^2 \beta^2 t^3 /(6\hbar M).$
Then, $ \hat{S}_x(t) = \hat{S}_z^2 \hat{S}_x \hat{S}_z^2$ which, with the help of the spin commutation relations, can also be written as $\hbar^2 \hat{S}_x \hat{S}_z^2 +2i\hbar \hat{S}_y \hat{S}_z^3 + \hat{S}_x \hat{S}_z^4$.
This expression shows in which sense this is a nonlinear spin transformation.

We assume that the initial state of the atoms is $\psi (\boldsymbol{r}) \sum_{m} c_{m} \ket{s,m},$ where $\psi (\boldsymbol{r})$ refers to the (spatial) wavefunction and $\sum_{m} c_{m} \ket{s,m},$  is an arbitrary spin pure state.
The spin state is characterized by its eigenvalues of the square spin and of the $z$-component of the spin, $\hat{\boldsymbol{S}}^2 \ket{s,m} = \hbar^2 s(s+1) \ket{s,m}$ and that $\hat{S}_z \ket{s,m} = \hbar m \ket{s,m},$ $s=0,1/2,1,\cdots$ $m = -s, -s+1, \cdots, s.$
From here on, we will drop the use of the label $s.$

When we apply $ \hat{U}_{2c}(t)$ to the initial state, we obtain
\begin{equation}
 [ \hat{U}_{2c}(t) \psi \sum_{m} c_{m} \ket{m}] (\boldsymbol{r}) = \sum_{m} c_{m}^{(2c)} (t) \psi\left(\boldsymbol{r} \right) \ket{m},
\end{equation} 
where $c_{m}^{(2c)}(t) = c_{m} \exp(-{i\hbar\gamma^2 \beta^2 m^2 t^3}/(6 M) ).$
We assume an initial Gaussian state which represents a particle moving with velocity $\boldsymbol{v} = v_0 \hat{\boldsymbol{\jmath}}$,
\begin{equation}
 \psi(\boldsymbol{r}) = N_0 \exp\left({-\frac{x^2}{4\sigma_x^2} -\frac{y^2}{4\sigma_y^2} -\frac{z^2}{4\sigma_z^2}+i\frac{M v_0 y}{\hbar}} \right),
\end{equation}
where $N_0 = 1/\sqrt{(2\pi)^{3/2} \sigma_x \sigma_y \sigma_z}$ is a normalization constant.
Here, $\sigma_x, \sigma_y$ and $\sigma_z$ represent the standard deviations along the coordinates $x, y$ and $z$, respectively.
Taking into account that momentum is the generator of spatial translations\cite{jordan1975why}
\begin{equation}
 [\exp(-i \boldsymbol{a}\cdot \hat{\boldsymbol{p}}/\hbar) \psi] (\boldsymbol{r}) = e^{- \boldsymbol{a}\cdot \hat{\boldsymbol{\nabla}}} \psi (\boldsymbol{r})
 = \psi(\boldsymbol{r} - \boldsymbol{a}),
\end{equation}
we see that $\hat{U}_{2b}(t) = \exp( -{i\gamma \beta t^2} \hat{S}_z \hat{p}_z/({2\hbar M}))$ produces a spin-dependent translation
\begin{equation}
 [ \hat{U}_{2b}(t) \psi \sum_{m} {c}_{m}^{(2c)} \ket{m}] (\boldsymbol{r}) = \sum_{m} {c}_{m}^{(2c)} \psi_m^{(2b)} (\boldsymbol{r}) \ket{m},
\end{equation}
where $\psi_m^{(2b)} (\boldsymbol{r}) =\psi(\boldsymbol{r} -\gamma \beta t^2 \hbar m \hat{\boldsymbol{k}}/(2 M) ). $ 
The spatial wavefunction is split into $2s+1$ parts, each associated with a different eigenvector of the $z$-component of the spin.
Each wavefunction component, $\psi_{m}^{(2b)}(\boldsymbol{r}),$ undergoes a different shift. 
Notably, there exists a connection that can be established with the semiclassical calculation, which predicts a force $F_z = \hbar m\gamma\beta,$ resulting in a (change of) momentum $F_z t =\hbar m \gamma\beta t  $ and a final deviation $Ft^2/(2M) = \gamma\beta m \hbar t^2/(2M).$
By integrating the semiclassical calculation with the knowledge of the form of the unitary operator for space translations, it becomes possible to infer the form of $\hat{U}_{2b}(t)$. 

The second term of the evolution operator, represented as $\hat{U}_{2a}(t) = \exp(-{i \hat{\boldsymbol{p}}^2 t}/(2M\hbar) ),$ corresponds to free evolution.
This particular scenario of a Gaussian wavefunction undergoing free evolution, often covered in introductory quantum mechanics courses, was considered by Blinder in a prior publication within this journal~\cite{blinder1968evolution}.
As time progresses, an initial Gaussian wavepacket centered at $r_0$, with an initial momentum of $p_0$, and a variance of $\sigma^2$ transforms into another Gaussian, centered at $r_0 + (p_0 t/M).$ 
Since the variance $\sigma^2$ is shifted by $i\hbar t/(2M)$ in the wavefunction, the physical position variance broadens over time, evolving as $\sigma^2 + \hbar^2 t^2/(4 M^2 \sigma^2).$ 
 
This analysis holds for the Gaussians along each of the three spatial components,
\begin{equation}
 [ \hat{U}_{2a}(t)  \sum_{m} {c}_{m}^{(2c)} \psi_{m} \ket{m}] (\boldsymbol{r}) = \sum_{m} {c}_{m}^{(2c)} {\psi}_{m}^{(2a)}(\boldsymbol{r}) \ket{m},
\end{equation} 
where
\begin{align} 
 {\psi}_{m}^{(2a)}&(\boldsymbol{r})   =  N(t) \exp\left( -\frac{(y-v_0 t)^2}{4\sigma_y^2(t)} +i\frac{M v_0 y}{\hbar}\right) \times \\
 &  \exp\left(-\frac{x^2}{4\sigma_x^2(t)}\right)  \exp\left( -\frac{(z- \gamma \beta t^2 \hbar m/(2 M))^2}{4\sigma_z^2(t)}  \right). \nonumber
\end{align}
Here, $ N(t) =  {1}/{\sqrt{(2\pi)^{3/2} \sigma_x (t) \sigma_y (t) \sigma_z (t)}}$, and $\sigma_i^2 (t) = \sigma_i^2 +i\hbar t/(2M),$ for $i=x,y,z$.
 
Finally, $\hat{U}_1(t)$  describes the position-dependent Larmor precession, that is, spin rotation around the $z$-axis, at frequency $\gamma (B_0 + \beta z) m $.
$\hat{U}_1(t)$  also represents a spin-dependent momentum kick $\gamma \beta \hbar t m$, because position is the generator of momentum translations. 
The spin translation described by $\hat{U}_{2b}(t)$ exactly corresponds to this shift in momentum.
We have
\begin{equation} \label{eq:finalstate}
 [ \hat{U}_1(t) \sum_{m} {c}^{(2c)}_{m} {\psi}^{(2a)}_{s} \ket{m}] (\boldsymbol{r}) = \sum_{m}  {c}^{(1)}_{m} {\psi}^{(1)}_{m}(\boldsymbol{r}) \ket{m},
\end{equation}
where  
\begin{eqnarray}
 {\psi}_{m}^{(1)}(\boldsymbol{r}) &=&\exp \left( i\gamma m t \beta {z}\right)  {\psi}_{m}^{(2a)},
\end{eqnarray}
 and ${c}^{(1)}_{m} = \exp \left( i\gamma m t B_0 \right)  {c}^{(2c)}_{m} $.
 
 $\hat{U}_{2a}(t)$ and $\hat{U}_{1}(t)$ are the evolution operators for their respective Hamiltonians, $\hat{H}_2$ and $\hat{H}_1$.
 In this sense, their appearance in the expression of the evolution operator of the whole Hamiltonian could have been anticipated.
 These operators take the form of exponentials of functions linear in time. 
 Consequently, the expectation values associated with the pertinent operators, such as position and momentum for $\hat{U}_{2a}(t)$ and spin for $\hat{U}_{1}(t),$ faithfully mirror the corresponding classical dynamics, depicting uniform motion and Larmor precession, respectively. 
 Based on the terms of the Hamiltonian and on the knowledge of the effects of unitary operators on the wavefunction and on other operators, it is possible to connect the semiclassical approach to the SGE with the quantum evolution operator, with the exception of a nonlinear spin transformation. 

It's important to note that under the experimental conditions of the Stern-Gerlach experiment (SGE), the wavefunction's broadening is minimal. 
Therefore, if the sole focus of the discussion pertains to the visibility of the splitting, a semiclassical treatment is entirely adequate.
Nonetheless, it's crucial to note that there is no superposition of classical trajectories in this context. 
Only through a quantum treatment do we observe the emergence of superpositions of products involving spatial and spin states, as exemplified in \eqref{eq:finalstate}.
The atomic state \eqref{eq:finalstate}, which challenges the notion of separate upper and lower beams~\cite{hiddenvariables}, cannot be written as a product of a spacial state times a spin state.  
 Unitary operators involving a product of position or momentum operators and a spin operator have the capacity to produce this type of states, known as entangled states~\cite{nonseparability}.
 These operators are able to transform a product state into an entangled state.

Let us consider the bipartite state 
\begin{equation} \label{eq:Schmidt}
 \ket{\Psi} =\sum_{m} \alpha_{m} \ket{\psi_{m}} \ket{m},
\end{equation}
where $\braket{\psi_{m}|\psi_{n}} = \delta_{mn} = \braket{m|n}.$ 
The density operator associated to the state \eqref{eq:Schmidt} is $\rho = \sum_{m,m'} \alpha_{m} \alpha_{m'}^* \ket{\psi_{m}} \ket{m} \bra{\psi_{m'}} \bra{m'}.$
The reduced spatial state $ \rho_s$ is obtaining tracing out the spin degree of freedom, $S,$
\begin{eqnarray}
 \rho_s &=& \operatorname{Tr}_S \rho = \sum_{n} \bra{n} \rho \ket{n} \\ &=&
\sum_{n} \sum_{m,m'} \alpha_{m} \alpha_{m'}^* \ket{\psi_{m}} \braket{n|m} \bra{\psi_{m'}} \braket{m'|n} \\
&= &
  \sum_{m} |\alpha_{m}|^2 \ket{\psi_{m}} \bra{\psi_{m}}.
\end{eqnarray}
The 	entanglement entropy $S(\rho_s) = -	\operatorname{Tr}(\rho_s \log (\rho_s)) $ is an entanglement measure for bipartite pure states.\cite{bennett1996mixed}
For states written in the form \eqref{eq:finalstate}, the entanglement entropy can be written as $S(\rho_s) = - \sum_{i} |\alpha_i|^2 \log |\alpha_i|^2.$

The atomic state described by Eq. \eqref{eq:finalstate} typically doesn't align with this particular form because the spatial states involved are not orthogonal. 
Nevertheless, there are two limiting cases in which it coincides with the representation in Eq. \eqref{eq:Schmidt}.
The first scenario occurs when the components $\ket{\psi_i}$ have not separated at all, essentially meaning that $\ket{\psi_i} = \ket{\psi}$. 
In this case, the state becomes factorized and conforms to the form in Eq. \eqref{eq:Schmidt}, featuring only one nonzero coefficient, $\alpha_1 = 1$.
As a direct consequence of this factorization, the entanglement entropy of the atomic state vanishes ($S(\rho_s)=0$). 
This result is entirely consistent with the description of a factorized state.
In the second scenario, when the spatial states associated with different values of the $z$-component of the spin are so neatly resolved that they can be considered orthonormal, the atomic state described in Eq. \eqref{eq:finalstate} already conforms to the form in Eq. \eqref{eq:Schmidt}. 
In this case,  $\alpha_m = c_m$, and the entanglement entropy is given by $S(\rho_s) = -\sum_{m} |c_{m}|^2 \log |c_{m}|^2.$
If, additionally, $|c_{m}| =1/\sqrt{2s+1},$ the entanglement entropy reaches its maximum 
\begin{equation}
 S(\rho_s) = -\sum_{m=-s}^s \frac{1}{2s+1} \log \frac{1}{2s+1} = \log(2s+1).
\end{equation}
To obtain this result, we used the fact that there are $2s+1$ terms in the sum.
We have found that, for a given pure initial state, the entanglement entropy of the atomic state increases with the quality of the measurement.
Here, the quality of measurement is defined as the degree of separation among the components of the wavefunction associated with distinct values of the $z$ component of the spin. 

Before concluding this paper, we would like to offer a brief comment on the probability density $p(\boldsymbol{r},t)$   to measure an atom at a point $\boldsymbol{r}$ at time $t$.
This probability is given by
\begin{eqnarray}
 p(\boldsymbol{r},t) &=& \sum_{m'} \bar{c}_{m'}^* \bar{\psi}_{m'}^*(\boldsymbol{r}) \bra{m'} \sum_{m} \bar{c}_{m} \bar{\psi}_{m}(\boldsymbol{r}) \ket{m}
\\ & = & \sum_{m} |c_{m}|^2 | \bar{\psi}_{m}(\boldsymbol{r})|^2,
\end{eqnarray}
where we used the orthonormality relation $\braket{m'|m}=\delta_{m,m'}.$ 
There are no interference between wavefunction components associated with different values of $\hat{S}_z.$
Thus, this probability density is the same for an initial pure spin state $\sum_{m} c_{m} \ket{m}$ and for an initial mixed operator $\sum_{m} |c_{m}|^2  \ket{m}\bra{m}.$\cite{note}
In particular, the calculated probability density for an initial state $(\ket{1/2}+\ket{-1/2})/\sqrt{2}$ coincides with that of the initial spin state comprising half of the atoms with spin up and half of them with spin down.
This mixed state describe the spin distribution in the original SGE.

In summary, while the customary semiclassical treatment of the Stern-Gerlach experiment is adequate for addressing aspects such as spin values and observed beam splitting, it falls short in conveying the fundamental concept of the superposition of trajectories. 
Only a quantum approach can accurately depict the superposition of spatial wavefunctions associated with distinct values of the $z$ component of spin.
We have introduced two simple methods for calculating the evolution operator and the atomic state, designed to be readily integrated into classroom instruction. 
Furthermore, our paper has offered an interpretation of the resulting evolution operator, emphasizing its connection with the semiclassical approach when applicable.
By embracing a quantum perspective, it is possible to deepen the discussion of the measurement process and to explore phenomena like entanglement and non-locality. 
Students can further investigate an interferometric setup where the gradient undergoes a sequence of changes from positive to negative and back to positive. 
The corresponding time intervals for these gradient changes are denoted as $T$, $2T$, and $T$ respectively.

\appendix*   

\section{Derivation of the operator expansion}

The Taylor series corresponding to $F(x)=e^{x \hat{A}} \hat{B} e^{-x \hat{A}}$ is
\begin{equation}
 F(x)= F(0) + \frac{x}{1!} F'[0]  + \frac{x^2}{2!} F''[0] + \cdots 
\end{equation}
Taking into account
\begin{equation}
  F'(x) = \hat{A} e^{x \hat{A}} \hat{B} e^{-x \hat{A}} - e^{x \hat{A}} \hat{B} e^{-x \hat{A}} \hat{A} =[\hat{A}, F(x)]
\end{equation}
we can calculate any derivative ($F^{(n+1)}(x) = [\hat{A},F^{(n)}(x)]$). Finally, since $F(0)=\hat{B}$, we obtain the operator expansion
\begin{equation}
 e^{x \hat{A}} \hat{B} e^{-x \hat{A}} = \hat{B}  + \frac{x}{1!} [\hat{A}, \hat{B}]   + \frac{x^2}{2!} [\hat{A}, [\hat{A}, \hat{B}]]  + \cdots 
\end{equation}

\begin{acknowledgments}


The author has no conflicts to disclose.

\end{acknowledgments}


\begin{thebibliography}{99}
\bibitem{spacequantization} The old quantum theory predicted that the elliptical orbits of Bohr and Sommerfeld could adopt only certain spatial orientations.
\bibitem{hiddenvariables} The pilot-wave interpretation reintroduces particles with definite positions. Within this interpretation it makes sense to talk about an upper beam and a lower beam.
\bibitem{nonseparability} 
The term entanglement is specifically reserved for denoting quantum correlations between degrees of freedom in spatially separated systems. In contrast, when these correlations involve degrees of freedom within a single system, the term nonseparability is employed.
\bibitem{note}
If only one of the coefficients $c_{m}$ does not vanish, this density operator represents a pure state.
Otherwise, it represents a mixed state.
\end{thebibliography}
\end{document}